%% file: Ulasmanuscript.tex
\documentclass[authoryear,preprint,review,12pt]{elsarticle}

\usepackage{graphics}
\usepackage{graphicx}
\usepackage{epsfig}
\usepackage{caption}
\usepackage{amssymb}
\usepackage{amsthm}
\usepackage{multirow}
\usepackage{amsmath}
\usepackage[flushleft]{threeparttable}

\def\astrobj#1{#1}
\journal{New Astronomy}

\begin{document}
\sloppy
\begin{frontmatter}


\title{First Multicolor Photometry of 1SWASP~J130111.22+420214.0 and 1SWASP~J231839.72+352848.2}

\author[abc]{B. Ula\c{s}\corref{dip}}
\author[xyz]{R. Michel}
\author[xyz]{S. Silva}
\address[abc]{\.{I}zmir Turk College Planetarium, 8019/21 sok., No: 22, \.{I}zmir, Turkey}
\address[xyz]{Observatorio Astron\'{o}mico Nacional, Instituto de Astronom\'{i}a, Universidad Nacional Aut\'onoma de M\'exico, Apartado Postal 877, 22830, Ensenada, B.C., M\'exico}
\cortext[dip]{Corresponding author \\
E-mail address: bulash@gmail.com}

\begin{abstract}
The first multicolor observations and light curve solutions of the stars 1SWASP~J130111.22+420214.0 and 1SWASP~J231839.72+352848.2 are presented.
The estimated physical parameters of 1SWASP~J130111.22+420214.0 are $M_1=1.0M_{\odot}$, $M_2=1.63M_{\odot}$, $R_1=0.77R_{\odot}$, $R_2=0.91R_{\odot}$, $L_1=0.59L_{\odot}$, and $L_2=0.68L_{\odot}$. While a binary model could not be fitted to the light curve of 1SWASP~J231839.72+352848.2, its Fourier series representation indicates that this system could well be a rotating ellipsoidal variable. The components of 1SWASP~J130111.22+420214.0 are also compared to other well--known binary systems and their position in mass--radius plane and the Hertzsprung--Russel diagram are argued and possible evolutionary statuses are discussed.
\end{abstract}


\end{frontmatter}

\section{Introduction}



Studies of binary star systems play an important role in filling some gaps in the understanding of stellar evolution, especially under the effects of binarity. Contact binaries, in this regard, provide valuable information for investigating the stellar evolution in the common envelope phase. 1SWASP~J130111.22+420214.0 (LINEAR~8209250, 2MASS~J13011120+4202138, SDSS~J130111.21+420213.8), hereinafter J1301, is a member of the huge group of detected contact binaries. It has a reported $V$ magnitude between 15.69 and 15.34. \cite{loh13} included J1301 in their list of 143 Super WASP eclipsing binaries and reported its period to be 19477.594 seconds (0.225435116 days). \cite{pal13} who cataloged the periods and several magnitude values for LINEAR variables, listed an orbital period of 0.225435 days for this object. Apart from these, there are no detailed studies of this system so far, possibly because it is a relatively faint object.

1SWASP~J231839.72+352848.2 (CRTS~J231839.7+352848, 2MASS~J23183973+3528479), hereinafter J2318, is listed as a binary system by \cite{loh13} and as an ultra--short period ellipsoidal binary by \cite{dra14}. As forJ1301, there is a lack of detailed studies on this target in the literature.

In the following sections, we give the details of our observations, explain the analyses of the light curves and then conclude with our results and a comparison of J1301 with other binary systems having similar configurations.

\section{Observations and reductions}
\label{obs}
Observations of both targets were made with the 0.84-m f/15 Ritchey-Chr\'etien telescope at San Pedro Martir Observatory (SPMO) attached with the Mexman filter-wheel and the \textit{Spectral Instruments} CCD detector (a 13.5 $\mu$m square 2048$\times$2048 deep depletion e2v CCD4240 chip with a gain of 1.32 e$^-$/ADU and a readout noise of 3.4 e$^-$). Full frame, with 2$\times$2 binning, was used during all the observations giving a field of view of $7.6^{\prime}\times7.6^{\prime}$.

J1301 was observed on March 24 2016 for a total of 8.8 hours of alternated exposures in filters $B$, $V$ and $R$ with exposure times of 60, 30 and 20 seconds respectively. A total of 481 images were acquired along with flat field and bias frames. All the target images were corrected by bias and flat field before the instrumental magnitudes of the stars were calculated with the standard aperture photometry method implemented in the IRAF\footnote{IRAF is distributed by the National Optical Observatories, operated by the Association of Universities for Research in Astronomy, Inc., under cooperative agreement with the National Science Foundation.} package. Field star 2MASS~J13011406+4157531 ($U=14.759\pm0.007$, $B=14.519\pm0.005$, $V=13.746\pm0.004$, $R=13.312\pm0.004$, and $I=12.879\pm0.004$ according to our measurements) was used as reference star to convert to relative magnitudes since it has similar colors to the variable.

J2318 was also observed, in the same manner (same filters and exposure times), in 2016 during the four nights of September 8, September 26, September 28 and October 21. Alternated exposures were taken during 5.4, 7.1, 3 and 2.3 hours respectively. Field star 2MASS~J23183579+3531184 ($U=17.334\pm0.009$, $B=16.616\pm0.001$, $V=15.660\pm0.001$, $R=15.140\pm0.003$, and $I=14.707\pm0.003$ according to our measurements) was used as reference star to convert to relative magnitudes.

The shapes of the light curves of both systems show typical W~UMa type variation with a slight magnitude difference between the primary and secondary minima. In addition, we calculated eight times of minimum light for the systems which are listed in Table \ref{tablist}.

\section{Light Curve Solutions}
\subsection{J1301}
\label{lc1}
\begin{figure}
\begin{center}
\includegraphics[scale=1.0]{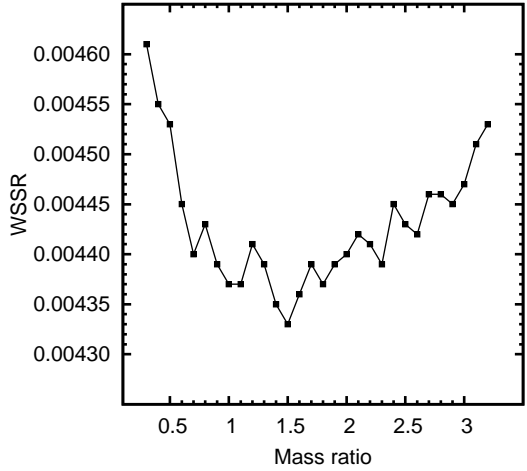}
\caption{The result of the q--search. WSSR stands for weighted sum of squared residuals.}
\label{figq}
\end{center}
\end{figure}

Since there is not any reported spectroscopic mass ratio for the system, we first applied the q--search method to find the best initial value to be used for $q$ during the light curve analysis. The mass ratio values were fixed from 0.3 to 3.2 with the step-size of 0.1 in simultaneous solution and the lowest residual value was found to be near $q=1.5$. The results are plotted in Fig.~\ref{figq}. We define the primary component as the less massive star and the secondary component as the more massive one, therefore, the mass ratio equation can be written as $q=M_2 / M_1$. 

The light curve solution was applied by using 2015 version of DC code\footnote{ftp://ftp.astro.ufl.edu/pub/wilson/lcdc2015/} which is based on the Wilson--Devinney method \citep{wil71}. The code applies best fits using differential corrections to yield the final parameters. The number of data points is 131 in $B$, 161 in $V$ and 160 in $R$ filter. The contact binary running mode was selected since the light curves show W~UMa type variations as mentioned in Sec.~\ref{obs}. The initial mass ratio value was set to 1.5 following the result of the q--search while the following parameters were left as free: inclination $i$, mass ratio $q$,  temperature of the secondary component $T_2$, surface potential $\Omega_1 = \Omega_2$ and luminosity of the primary component $L_1$. The albedo values $A_1$, $A_2$ were adopted from \cite{ruc69} and the gravitational darkening coefficients $g_1$, $g_2$ were calculated as in \cite{ham93}. The effective temperature for the primary component was determined by a calibration process. We first derived the $E(B-V)$ for the system as 0.11, then we calculated the intrinsic $B-V$ color and estimated the temperature value by using the correlations given by \cite{flo96}. The very slight difference between two maxima in the $R$-filter directed us to give a trial with a cool spotted model approximation, however, we could not achieve physically meaningful results fitting with a spotted area in the simultaneous solution of the light curves.

\begin{figure}
\begin{center}
\includegraphics[scale=1.0]{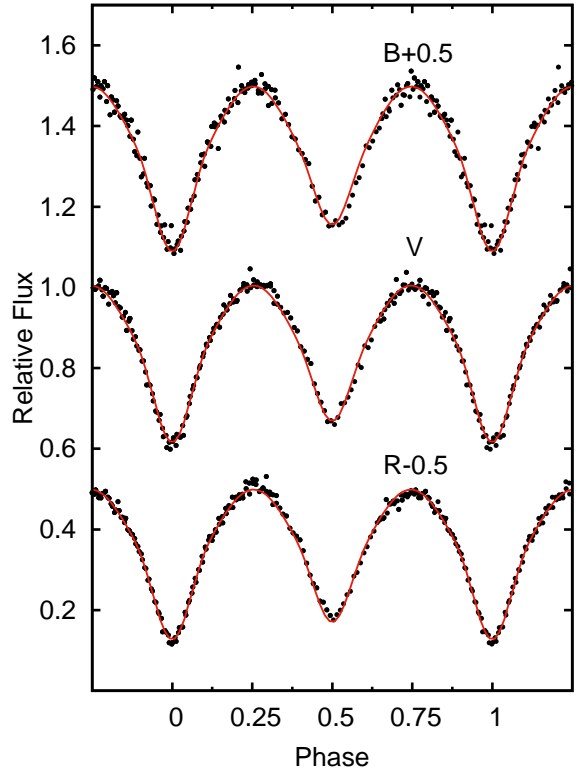}
\caption{The calculated (lines) and the observational (dots) light curves of J1301. $B$ and $R$ curves are shifted for the sake of visibility.}
\label{figlc}
\end{center}
\end{figure}

\begin{figure}
\begin{center}
\includegraphics[scale=1.0]{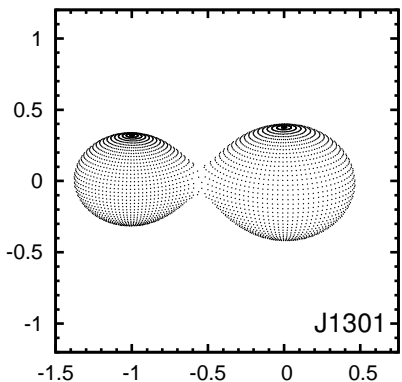}
\caption{The estimated geometry of J1301 at $\Phi=0.75$}
\label{figgeo}
\end{center}
\end{figure}

The present solution of the light curve is the first analysis of the system in the literature and gives the results which fit with a typical W--type W~UMa system that the massive component has the lower temperature value. The primary component (hotter) contributing about 45\% to the total light in average. The results of the solution are listed in the second column of Table~\ref{tablc1} and plotted in Fig.~\ref{figlc}. The estimated geometry of the system at 0.75 phase is also drawn in Fig.~\ref{figgeo}.

\subsection{J2318}
\subsubsection{Binarity}
\label{bin2}
Since J2318 was classified as an eclipsing binary candidate by \cite{loh13}, we first solved the light curve of the star with a binary system assumption. Since the shape of the light curve indicates a contact binary, we used  Mode 3 for "overcontact binary not in thermal contact" mode of DC code (see Sec.~\ref{lc1}) to achieve the best solution.  

We followed the same procedure stated in Sec.~\ref{lc1} during the analysis. The number of the observational points was 278, 279 and 278 in $B$, $V$, and $R$ filters, respectively. The temperature of the primary component was assumed as 4930~K according to the derived E(B-V) value. First, a q--search applied to the light curves since no spectroscopic mass ratio available in the literature. $q$ value from 0.2 to 1.0 was tried during the investigation, however, the solutions gave a potential surface value smaller then the $L_2$ potential which means that the configuration is conflict with our hypothesis that the system is a contact binary. Therefore, a physically meaningful solution could not be reached and we could not proceed to the light curve analysis without a mass ratio value supported by a solution with contact binary configuration.

The unsuccessful attempts of solving the light curve by assuming that the system is contact binary empowered the possibility that the star is an ellipsoidal variable which we mention the details in the next subsection. 
 
\subsubsection{Ellipticity}
We investigated the ellipticity effect in the light of J2318 since $\it{(i)}$ the system was classified as ellipsoidal variable in the catalog of \cite{dra14} and $\it{(ii)}$ the results of the binary solution did not allow us to construct a physically meaningful binary system (Sec.~\ref{bin2}). The light variation is assumed to occur because of the oblate shape of the components rather than eclipses of the components. \cite{hal90} mentioned that $\cos(2\theta)$ term of the representative Fourier series may refer to ellipticity and chromospheric activity (two dark regions). This two effects can be distinguished by inspecting the properties of the light variation. The ellipticity effect shows constant amplitude and minimum at two conjunctions while the amplitude of the light variation rising from the stellar spot varies in one year, a very different value from the orbital period. 

\begin{figure}
\begin{center}
\includegraphics[scale=1.0]{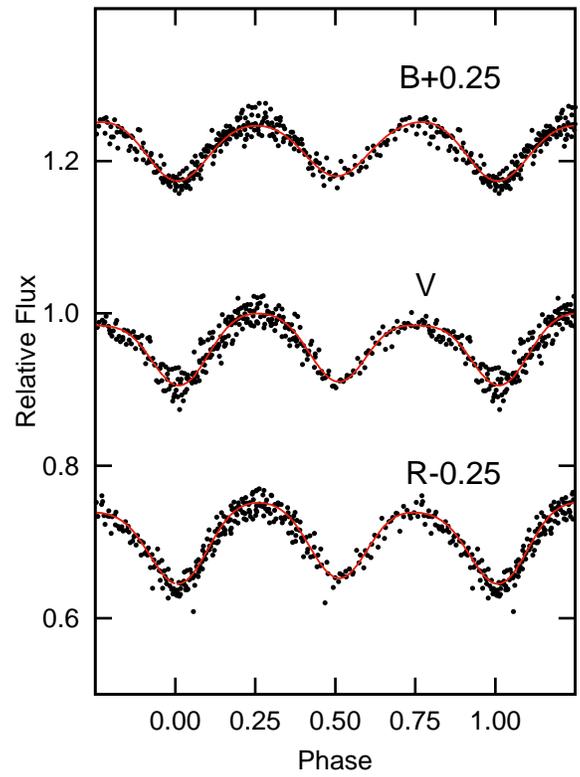}
\caption{Agreement between observations (dots) and the calculated light curves for J2318 by assuming that the ellipticity effect is responsible for the light variation. $B$ and $R$ curves are shifted for the sake of visibility.}
\label{figfou}
\end{center}
\end{figure}

For investigating the ellipticity we inspect the dominance of the coefficients by representing the light curve with a Fourier series of the form:

\begin{align*}
L(\theta)=\frac{a_0}{2} + \sum\limits_{k=1}^n (a_k \cdot \cos({k\theta}) + b_k \cdot \sin({k\theta}))
\end{align*}
where  $a_0=\frac{1}{\pi}{\displaystyle \int_{-\pi}^{\pi}}L(\theta)d\theta$, $a_k=\frac{1}{\pi}{\displaystyle\int_{-\pi}^{\pi}} L(\theta) \cos(k\theta)d\theta$, $b_k=\frac{1}{\pi}{\displaystyle\int_{-\pi}^{\pi}} L(\theta) \sin(k\theta)d\theta$. The theoretical light curve obtained from the above equation, along with the observations, are plotted in Fig.~\ref{figfou}. The $a$ and $b$ coefficients involved with the $\cos$ and $\sin$ terms, respectively are listed in Table~\ref{tabfr_2}. It can be seen that the $a_2$ coefficients, regarding the $\cos(2\theta)$ terms, are dominant in all filters. Moreover, the $a_1$ term which corresponds to one dark region \citep{hal90} was also not found large enough to assign a single spotted area as the origin of the observed light variation. 

We conclude that the light variations are most likely a consequence of the ellipticity. However, future observations spread through one year are needed to eliminate, or confirm, any effect from chromospheric active areas.

\section{Discussion and Conclusions}

We presented the first multicolor light curve solution of the contact binary system J1301 and the possible ellipsoidal variable J2318. New times of minimum light are given and light parameters are provided. The absolute parameters were estimated by assuming that the primary component of J1301 is a main--sequence star and using the (B-V)--mass calibrations given by \cite{cox00} (Table~\ref{tababs1}). The fill--out factor of J1301 were also calculated by using the formula given by \cite{luc79}:

\begin{align*}
f=\frac{\Omega_{i}-\Omega}{\Omega_{i}-\Omega_{o}}
\end{align*}
where $\Omega_{i}$ and $\Omega_{o}$ refers the inner and the outer Lagrangian potentials. The factor for the system is found to be 16\%.

It was found that J1301 is an W--type W~UMa system, in agreement with the classification of \cite{bin70} who proposed two sub--classes for these type of stars: A--type systems show light curves where the more massive component is eclipsed during the primary (deeper) minimum while the secondary (shallower) minimum corresponds to occultation of the less massive component. On the other hand, the less massive component is eclipsed during the primary (deeper) minimum of the light curves of W--type systems where the occultation of the more massive component takes place during the secondary (shallower) minimum. \cite{mac96} suggested that the A-- and W-- type systems are members of cool contact binaries. They also propose that the W--type binaries could evolve to A--types while A--type systems stay stable during contact configuration through their evolution. The evolution of contact binaries was also studied by \cite{ste12} who remarked that as the massive star of the detached binaries evolves, it fills its Roche lobe and starts the mass transfer to the less massive component. The mass transfer ends shortly after the mass ratio reverses and angular momentum loss supports the formation of the contact system. The scenario may end with the merging of the components generating a rapidly rotating star.

\begin{figure}
\begin{center}
\includegraphics[scale=1.0]{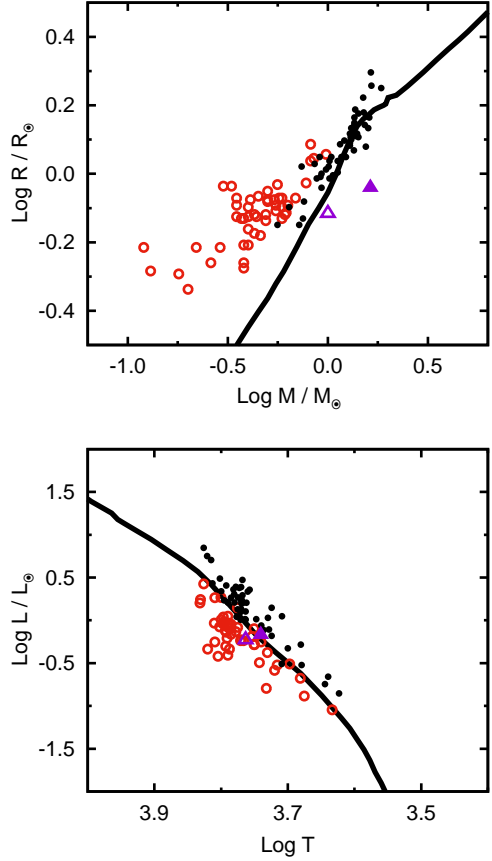}
\caption{The location of the components of J1301 in M--R plane (upper panel) and the Hertzsprung--Russel diagram (lower panel) . The components of our targets are shown with triangles. Filled and open signs refer to the primary and the secondary components. The data for W-type contact binaries are taken from \cite{yil13}. Note that we reversed the masses of the components since the authors mentioned that the more massive star set as primary in their catalog. The solid line in the H--R diagram indicates the ZAMS whose data are taken from \cite{gir00}. See text for details.}
\label{fighrmr}
\end{center}
\end{figure}

A comparison of J1301 to other 49 well--known W--type contact binaries \citep{yil13} can be found in Fig.~\ref{fighrmr}. Since the authors referred the more massive components as primaries in their paper we reversed our mass ratio value to be able to compare our final results to other systems. In the H--R diagram the secondary component is located under the ZAMS as the primary was hypothesized as for the main sequence star. The components also seems to have smaller radius according to its mass when compared to other contact binaries in mass--radius plane. Since a simultaneous solution covering the radial velocity data gives more precise results a velocity curve to be yielded from the spectroscopic observations of the system are needed for more precise determination of the physical parameters.

Based on our observations, we hypothesize that the light variation of J2318 is a consequences of eclipses in a binary system, however, we could not verify it by using analyzing methods for binary systems. Therefore, we concluded that the system is not a contact binary as we mentioned in our hypothesis and ellipticity effects are more likely to be involved with the observed light variation of the target.



\section*{Acknowledgments}
The authors would like to thank the anonymous referee for his/her valuable comments and suggestions. This paper is based on observations acquired at the OAN-SPM, Baja California, Mexico. RM acknowledge the financial support from the UNAM under DGAPA grant PAPIIT IN 105115.

\newpage

\input{Ulastab1.tex}
\input{Ulastab2.tex}
\input{Ulastab3.tex}
\input{Ulastab4.tex}

\end{document}

%% file: Ulastab1.tex
\begin{table}
\begin{center}
\caption{New times of minimum light calculated from our observations.}
\renewcommand{\arraystretch}{1.1}
\label{tablist}
\begin{tabular}{lcc}
\hline
System & Min type		&Value		 \\
\hline
J1301&I	&2457471.7386(3) \\
&II	&2457471.8530(5) \\
&I	&2457471.9642(1) \\
J2318 &II &2457639.846(1) \\
  &I &2457639.9495(6)  \\
  &II &2457657.763(2)  \\
  &I &2457657.8616(8)  \\
  &I &2457658.8690(8)  \\
\hline
\end{tabular}
\end{center}
\end{table}

%% file: Ulastab2.tex
\begin{table}
\begin{center}
\footnotesize
\caption{Results of the light curve analysis of J1301. Formal error estimates are given in parenthesis.}
\renewcommand{\arraystretch}{1.1}
\label{tablc1}
\begin{tabular}{lc}
\hline
Parameter      		    &  J1301   \\
\hline
$i$~${({^\circ})}$ 	   	&71.3(2)	\\
$q$		   	     	&1.63(1)	\\
$T_{1}$~(K)	   	     	&5800		\\
$T_{2}$~(K)	   	     	&5501(54)       \\
$\Omega_{1}$    		&4.622(7)  	\\
$\Omega_{2}$ 			& $\Omega_{1}$  \\
$\overline{r_{1}}$	  	&0.348(1)  	\\
$\overline{r_{2}}$	        &0.432(2) 	\\
$f$ (\%)		     	&16 	 	\\
$\frac{L_{1}}{L_{1}+L_{2}}$:    & 	    	\\
$B$ 				& 0.473(8) 	\\ 
$V$ 				& 0.453(5) 	\\
$R$ 				& 0.441(5) 	\\
\hline
\end{tabular}
\end{center}
\end{table}

%% file: Ulastab3.tex
\begin{table}
\begin{center}
\caption{List of Fourier coefficients obtained from the analysis of the light curves of J2318. See text for details.}
\renewcommand{\arraystretch}{1.1}
\label{tabfr_2}
\begin{tabular}{lccc}
\hline
Coefficient   &  B & V & R  \\
\hline			    
$a_0$ 	  &~1.9336	& ~1.9149  & ~1.9038\\
$a_1$	  &-0.0011  & -0.0024 &-0.0039  \\
$a_2$	  &-0.0356  & -0.0417 &-0.0478   \\
$a_3$     &-0.0019  & -0.0001 & -0.0001   \\
$a_4$     &-0.0041  & -0.0072 & -0.0073   \\
\hline
$b_1$     & -0.0023 & -0.0057 & -0.0057 \\
$b_2$	  & -0.0035 & -0.0054 & -0.0053   \\
$b_3$	  &-0.0003  & -0.0021 & -0.0006   \\
$b_4$ 	  &~0.0002   & -0.0024 & -0.0023   \\
\hline
\end{tabular}
\end{center}
\end{table}

%% file: Ulastab4.tex
\begin{table}
\caption{Estimated absolute parameters of the system J1301. P and S letters stand for the primary (less massive) and the secondary (more massive) components. The effective temperature of the sun is set to 5780~K.}
\label{tababs1}
\begin{tabular}{lcc}
\hline 
&  \multicolumn{2}{c}{J1301}  \\
Parameter &   \multicolumn{1}{c}{P} &   \multicolumn{1}{c}{S}  \\
\hline
{\it Mass}~(M$_{\odot}$)  & 1.0 & 1.63(1)  \\
{\it Radius}~(R$_{\odot}$) & 0.766(5) & 0.910(5)  \\
{\it Effective Temperature}~(K) & 5800 &5501(54)   \\
{\it Luminosity} ~(L$_{\odot}$) &0.594(8) & 0.680(7)  \\
{\it Semi--major axis}~(R$_{\odot}$) &\multicolumn{2}{c}{2.202(5)} \\
\hline
\end{tabular}
\end{table}